\documentclass{article}

\usepackage{microtype}
\usepackage{graphicx}
\usepackage{subfigure}
\usepackage{booktabs}
\usepackage{hyperref}

\usepackage{mlsys2025}
\usepackage{amsmath,amsfonts,amssymb,xspace,xcolor,enumitem}
\usepackage{tabularx,array,multirow,makecell}
\usepackage{siunitx,float,placeins}

\graphicspath{{figures/}}
\newcommand{\sys}{\textsc{Tideway}\xspace}

\makeatletter
\gdef\isaccepted{1}
\renewcommand{\Notice@String}{Preprint. Under review.}
\makeatother

\hypersetup{
  pdfauthor={Yifei Wang, Tsun-yu Yang, Yechen Xu, Hancheng Ye, Qinsi Wang, Baijun Zhou, Tingjun Chen, Hai Helen Li, Danyang Zhuo, Yiran Chen},
  pdfsubject={Public research preprint},
  pdfkeywords={Mixture-of-Experts, MoE inference, phase-aware tiered memory, consumer devices},
}

\mlsystitlerunning{Tideway: Phase-Aware Tiered Memory Management for MoE Inference on Consumer Devices}

\begin{document}

\twocolumn[
\mlsystitle{Tideway: Phase-Aware Tiered Memory Management for \\ MoE Inference on Consumer Devices}

\begin{mlsysauthorlist}
  \mlsysauthor{Yifei Wang}{duke}\hspace{0.8em}
  \mlsysauthor{Tsun-yu Yang}{duke}\hspace{0.8em}
  \mlsysauthor{Yechen Xu}{duke}\hspace{0.8em}
  \mlsysauthor{Hancheng Ye}{duke}\hspace{0.8em}
  \mlsysauthor{Qinsi Wang}{duke}\\
  \mlsysauthor{Baijun Zhou}{duke}\hspace{0.8em}
  \mlsysauthor{Tingjun Chen}{duke}\hspace{0.8em}
  \mlsysauthor{Hai ``Helen'' Li}{duke}\hspace{0.8em}
  \mlsysauthor{Danyang Zhuo}{duke}\hspace{0.8em}
  \mlsysauthor{Yiran Chen}{duke}
\end{mlsysauthorlist}
\mlsysaffiliation{duke}{Duke University, Durham, North Carolina, USA}
\vskip 0.18in
\begin{abstract}

Large language model (LLM) agents are rapidly becoming a dominant paradigm for AI applications. Recently, there is a growing demand to deploy large-scale Mixture-of-Experts (MoE) models locally on consumer PCs to balance data privacy, API costs, and inference efficiency. However, unlike datacenter servers with abundant DRAM, consumer devices are tightly constrained in both GPU VRAM and host RAM, forcing massive MoE weights to spill across VRAM, host RAM, and SSD.
Under such spilling, we observe that existing systems fail to fully utilize this memory hierarchy, treating the tiers as independent caches and thrashing in inference.

We present \sys, a system for efficient inference of large MoE models on memory-constrained consumer devices. Our core insight is to manage host RAM and GPU VRAM as a unified memory pool. Within this pool, \sys couples phase-aware memory layouts, which adapt to the distinct access patterns of prefill and decode, with an execution engine that optimizes CUDA Graph execution under expert paging. Our evaluations on representative consumer hardware with frontier MoE models demonstrate that \sys significantly outperforms state-of-the-art systems. On average, \sys achieves speedups of 1.97$\times$ in end-to-end latency, 2.40$\times$ in Time-to-First-Token (TTFT), and 1.61$\times$ in Time-Per-Output-Token (TPOT) against the fastest baseline.

Our source code is publicly available at \url{https://github.com/Afterglow231/Tideway}.

\end{abstract}

\vskip 0.12in
]

% Affiliation appears below the first column, as in the reference preprint;
% no personal e-mail address is invented or disclosed here.
\footnotetext[1]{Duke University, Durham, North Carolina, USA.\par\smallskip Preprint. Under review.}
\hypersetup{pdfauthor={Yifei Wang, Tsun-yu Yang, Yechen Xu, Hancheng Ye, Qinsi Wang, Baijun Zhou, Tingjun Chen, Hai Helen Li, Danyang Zhuo, Yiran Chen}}

\section{Introduction}
\label{sec:intro}

Large language models (LLMs) increasingly serve as the control core of autonomous agents, now in everyday use for research~\cite{openaiDeepResearch2025}, computer use~\cite{xie2024osworld}, and, most visibly, software engineering~\cite{anthropic2025claudecode,openai2025codex}. These interactive agents are mostly powered by cloud APIs. Meanwhile, open-weight mixture-of-experts (MoE) models now offer strong reasoning and coding capability while activating only a fraction of their parameters per token~\cite{yang2025qwen3,deepseekai2026deepseekv4}, and self-hosting them brings data privacy, reduces the token costs from long-running tasks, and permits domain-specific adaptation~\cite{song2024powerinfer,erdogan2024tinyagent}. Consequently, users are increasingly motivated to run these agents locally and cheaply, keeping sensitive data on-device.

Recently, open-source engines such as FreeToken~\cite{yang2026freetoken} and llama.cpp~\cite{gerganov2023llamacpp} are bringing frontier-scale MoE models to personal PCs with consumer GPUs. However, even high-end consumer GPUs lack sufficient VRAM to hold the largest models, shifting much of the storage burden to host RAM.
Particularly, for many users, a capable GPU and 16--64\,GiB of system RAM are sufficient for everyday computing. However, upgrading to hundreds of gigabytes of RAM solely to accommodate large AI model weights can entail a substantial additional investment. Growing AI demand further increases this cost because GPU memory and system RAM compete for the same DRAM manufacturing capacity, driving up memory prices~\cite{trendforce2026memory}.
On such PCs, weights spill beyond VRAM into host RAM and, when the combined memory budget is insufficient, extend to SSD.

A central challenge is \textit{how to minimize costly weight transfers while sustaining efficient GPU execution under expert paging}. The activated experts are decided at runtime, so cross-tier transfers land on the critical path of every forward pass. This cost appears in both phases. During prefill, a long prompt activates nearly every expert of every layer, so the whole model streams into VRAM before the first token is produced, inflating Time-to-First-Token (TTFT). During decode, each step touches only a few experts, but any missing expert stalls the decoding step and lengthens Time-Per-Output-Token (TPOT). Across long generations and successive turns, these stalls accumulate into end-to-end latency.
Expert layout across tiers thus decides not only whether the model fits but also its inference performance.

Existing inference systems struggle with this challenge, as their memory assumptions and optimization strategies do not match a consumer PC. FreeToken optimizes prefill and decode separately but assumes the complete expert pool resides in host RAM; its VRAM cache stores copies of experts already held in RAM, and the design has no fallback once the pool exceeds host memory. llama.cpp offers offloading and hybrid CPU--GPU execution but delegates paging to the OS through memory-mapped files, which operate at page rather than expert granularity and thrash once the working set reaches SSD. FlashMoE~\cite{kim2026flashmoe} supports SSD-backed inference, but its learned replacement policy still thrashes under full-model prefill scans. SSD-LLaMA~\cite{liang2026ssdllama} cuts expert-loading overhead with compression, asynchronous transfers, and cross-tier caching, yet keeps one layout fixed across phases, leaving memory underutilized and incurring avoidable transfers.
Ultimately, \textit{none of these systems treats the constrained GPU and host memory as a unified pool, failing to utilize the full potential of the memory hierarchy on consumer PCs}.

We design and implement \sys, a system for efficient large MoE inference on consumer PCs, which couples a phase-aware memory manager with an execution engine.
Instead of treating host RAM as a swap space behind VRAM, the manager combines GPU and host memory into a unified pool backed by the SSD. To maximize effective capacity and reduce expert misses, \sys deduplicates experts across the VRAM and RAM tiers and exploits bi-directional PCIe bandwidth, selectively returning evicted GPU experts to host RAM to avoid subsequent SSD reloads.

Within the pool, we introduce a phase-aware strategy.
\sys maintains separate memory layouts for streaming prefill and sparse decode, and switches between them at the phase boundary.
This adaptation preserves optimal memory states and actively prevents cross-tier thrashing during phase transitions.
To sustain efficient GPU execution as expert residency changes, \sys preserves CUDA Graph replay for reusable portions of decode computation.
The engine dynamically fetches out-of-core experts and retains graph-based execution wherever possible, limiting per-token host orchestration and kernel-launch overhead. Together, these mechanisms improve both prefill and decode performance without requiring the full expert pool to fit in host memory.

We evaluate \sys on an NVIDIA RTX~5090 with 32\,GB of VRAM under host RAM budgets of 16--128\,GiB, using three open-weight MoE models: Qwen3.6-35B-A3B~\cite{qwen2026qwen36}, Gemma4-26B-A4B~\cite{gemmateam2026gemma4}, and DeepSeek-V4-Flash. The evaluation covers 270 fixed-shape configurations with input lengths of 4K to 128K tokens and output lengths of 128 to 2K tokens, and trace-driven replays of eight multi-turn coding-agent sessions from TraceLab~\cite{zhu2026tracelab}. Across all fixed-shape configurations, \sys achieves 1.97$\times$ average and up to 3.18$\times$ end-to-end speedups over the fastest baseline, with 2.40$\times$ and 1.61$\times$ average gains in TTFT and TPOT. On TraceLab replays of coding-agent sessions, \sys further reduces total request processing time by up to 3.14$\times$.

Our paper makes the following contributions:
\begin{itemize}[topsep=3pt, itemsep=3pt, parsep=0pt, leftmargin=*]

    \item We characterize the performance degradation of existing MoE inference systems under limited host RAM, identifying SSD traffic and execution overhead as key bottlenecks.

    \item We design \sys, coupling a unified, phase-aware memory manager with an execution engine that preserves CUDA Graph replay under expert paging, reducing SSD accesses and per-token overhead.

    \item We evaluate \sys on frontier-scale MoE models. Against the fastest baselines, \sys achieves $1.97\times$ average end-to-end speedup on fixed-shape workloads and up to $3.14\times$ speedup on coding-agent traces.

\end{itemize}

\section{Motivation}
\label{sec:motivation}

\subsection{Memory on Consumer PCs}

Deploying a frontier-scale MoE model locally offers compelling privacy and cost benefits, but a consumer PC rarely has the memory to hold one. Four 2026 datasets, RTX~5090 desktops on sale from system integrators, 3DMark submissions~\cite{ul2026threedmark} and BOINC hosts~\cite{boinc2026hosts} with an RTX~5090, and the Steam Hardware Survey~\cite{valve2026steam}, tell the same story: 16--64\,GiB of RAM is the mainstream, covering 64--84\% of the three RTX~5090 populations and 99\% of Steam machines, while more than 128\,GiB appears in at most 11\%, with the largest share among BOINC hosts. Even owners of the top consumer GPU stop at 64\,GiB.

\begin{figure}[t]
    \centering
    \includegraphics[width=\linewidth]{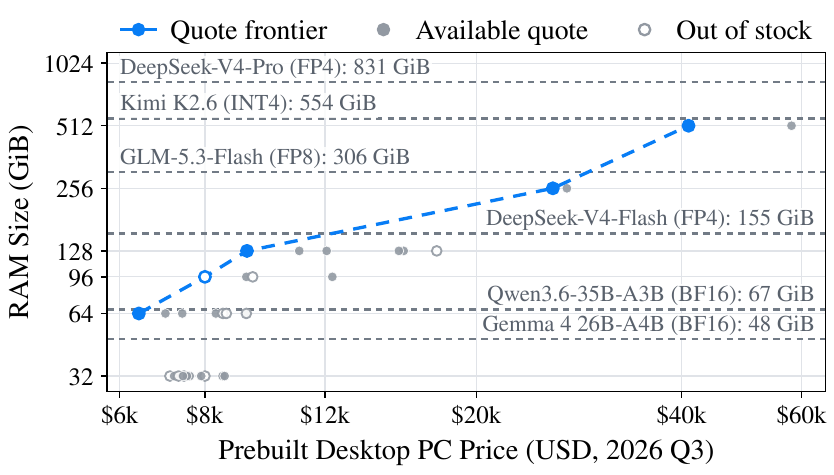}
    \vspace{-8mm}
    \caption{Prebuilt desktop quotes by price and RAM size (dots), and the weight size of open-weight MoE models (dashed lines).}
    \label{fig:ram_price}
    \vspace{-3mm}
\end{figure}

\begin{figure*}[t]
  \centering
  \includegraphics[width=\linewidth]{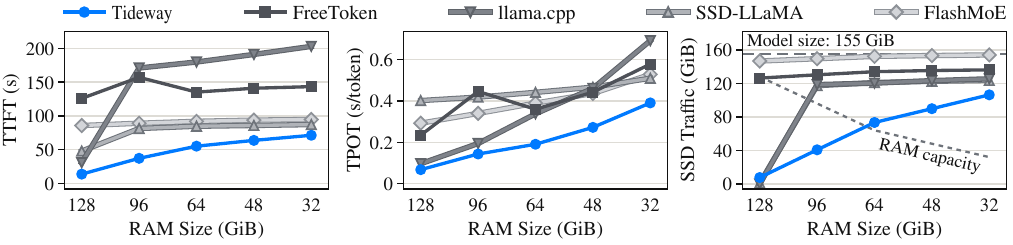}
  \vspace{-8mm}
  \caption{TTFT, TPOT, and SSD traffic of DeepSeek-V4-Flash-FP4 on an RTX~5090 as host RAM shrinks from 128 to 32\,GiB, replaying realistic agent traces in TraceLab with prefill chunk size set to 8K tokens.}
  \label{fig:ram-sweep}
  \vspace{-5mm}
\end{figure*}

\begin{figure}[t]
    \centering
    \includegraphics[width=\linewidth]{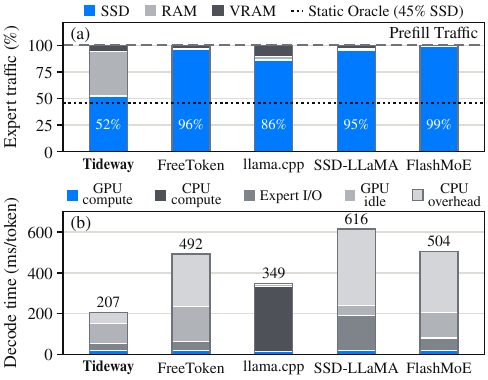}
    \vspace{-8mm}
    \caption{(a) Prefill expert traffic and (b) breakdown of per-token decode time for DeepSeek-V4-Flash on an RTX~5090 with 64\,GiB of host RAM (32K-token prompt, 1K output tokens).}
    \label{fig:pd}
    \vspace{-3mm}
\end{figure}

Purchasing a desktop with substantially more RAM requires a different system budget. Each point in Figure~\ref{fig:ram_price} is a prebuilt desktop quote plotted by price and RAM size, and the blue line traces the cheapest quote at each capacity. The frontier stays under \$10k up to 128\,GiB and then bends sharply, to about \$26k at 256\,GiB and \$41k at 512\,GiB. The dashed lines mark open-weight MoE models, which show that even expensive configurations cannot accommodate the full range of frontier MoE models.
Provisioning sufficient RAM for full model residency is therefore a costly requirement, not a realistic default for consumer devices.

\subsection{MoE Inference with Insufficient RAM}
\label{sec:motivation:moe}

When GPU and host memory cannot retain the full model, loading missing experts from SSD becomes part of the inference path.
Figure~\ref{fig:ram-sweep} quantifies this cost by replaying agent traces with DeepSeek-V4-Flash-FP4 on an RTX~5090 using 8K-token prefill chunks.
Once weights spill to SSD, baselines read most of the 155\,GiB model from storage regardless of host RAM size, keeping TTFT persistently high.
Across the full sweep from 128 to 32\,GiB, baseline TTFT and TPOT increase by factors of approximately $1.1\times$--$6.4\times$ and $1.3\times$--$7.0\times$, respectively.
In contrast, our design coordinates cross-tier residency and execution, enabling \sys to cut SSD traffic and achieve lower TTFT and TPOT across all RAM budgets during both phases.

This performance degradation reflects limitations in both expert retention and execution. These systems fail to keep useful experts resident, and they execute inefficiently while paging. Figure~\ref{fig:pd} examines these limitations using a 32K prompt and 1K output tokens with 64\,GiB of host RAM.

During prefill, each 8K-token chunk activates nearly the entire expert pool, requiring weights to stream through the GPU once per chunk.
Existing systems were built on the assumption that host RAM is a backing cache of the VRAM. When the pool exceeds RAM, simply extending this hierarchy with LRU-style demand paging fails to preserve expert reuse across chunks. Under a cyclic scan, loading later-layer experts evicts earlier-layer experts, causing successive chunks to repeatedly reload the same weights from SSD.
Figure~\ref{fig:pd}(a) shows that SSD supplies 86--99\% of the baselines' prefill expert traffic, although the 64\,GiB RAM holds 44\% of the pool. \sys keeps resident experts in place across the scan, and fetches only 52\% from SSD.

Decode exposes a distinct execution bottleneck beyond weight I/O.
CUDA Graphs typically reduce kernel launch overhead by replaying a captured, fixed-shape sequence, but that fixed shape complicates end-to-end capture with dynamic cross-tier paging. Existing systems thus fall back to eager execution for expert miss handling, increasing host dispatch and synchronization overhead.
Figure~\ref{fig:pd}(b) shows that CPU orchestration overhead and GPU stalls dominate decode time in most baselines, whereas llama.cpp incurs little launch overhead but spends 92\% of decode time computing experts on the CPU.
\sys minimizes CUDA Graph disruptions under expert paging, reduces expert I/O and CPU overhead, and achieves lower overall decode latency than all baselines.

In summary, accelerating MoE inference on consumer PCs requires making effective use of the memory capacity, while sustaining efficient execution across inference phases.

\newcommand*\circled[1]{\tikz[baseline=(char.base)]{
  \node[shape=circle,draw,inner sep=1pt,fill=white,thick] (char)
  {\footnotesize\textbf #1};}}

\section{Overview}
\label{sec:arch_overview}

\begin{figure}[t]
    \centering
    \includegraphics[width=\linewidth]{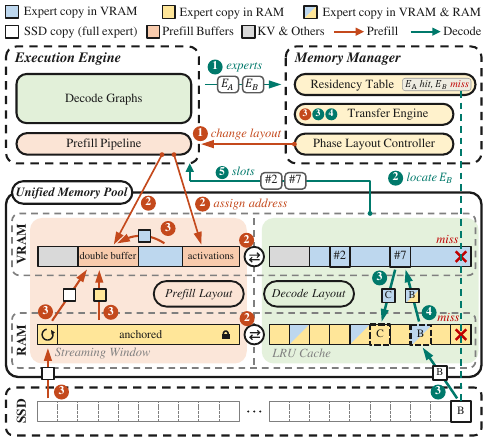}
    \vspace{-6mm}
    \caption{System architecture and phase-aware execution workflows of \sys across the three-tier memory hierarchy.}
    \label{fig:overview}
    \vspace{-4mm}
\end{figure}

Motivated by the findings in \S\ref{sec:motivation}, \sys{} is designed to enable  frontier-scale MoE model inference on consumer PCs where model weights inevitably spill across VRAM, host RAM, and SSD. To overcome the structural bottlenecks of existing inference systems, \sys pursues two core design goals: (1)~\emph{maximizing effective memory capacity} across VRAM and host RAM while dynamically adapting memory layouts to prevent cross-tier thrashing, and (2)~\emph{sustaining efficient GPU execution} across both prefill and decode phases despite high-latency SSD paging.

Figure~\ref{fig:overview} presents the overall architecture of \sys. Across the underlying storage hierarchy, \sys abstracts GPU VRAM and host RAM as a \emph{Unified Memory Pool}, while treating the SSD as an out-of-core backing store for spilled weights.
Atop this hierarchy, \sys{} introduces two co-designed components: an \emph{Execution Engine} and a \emph{Memory Manager}. To regulate expert residency and cross-tier data movement, the memory manager tracks expert locations through the \emph{Residency Table}, reconfigures pool partitions via \emph{Phase Layout} controller, and dispatches asynchronous migrations through the \emph{Transfer Engine}. Concurrently, the execution engine drives model computation through phase-specialized execution paths, pairing a streaming \emph{Prefill Pipeline} with paging-aware \emph{Decode Graphs}. Together, these two components coordinate memory layout adaptation and kernel execution across inference phases.

\paragraph{Prefill path.}
During prefill (red arrows), every chunk activates nearly all experts, so the memory pool is arranged for streaming rather than caching. The Phase Layout Controller \textcircled{\scriptsize 1} first directs the Execution Engine to switch to the \emph{Prefill Layout}. The switch \textcircled{\scriptsize 2} reshapes both tiers: in VRAM, the Prefill Pipeline carves out an activation region and a double buffer for weight staging, borrowing the space from the decode expert cache; in RAM, the resident experts are anchored against eviction for reuse across chunks, and a small streaming window is reserved for experts arriving from SSD. The Transfer Engine \textcircled{\scriptsize 3} then fills the double buffer alternately with expert weights from VRAM, RAM, or SSD, overlapping weight delivery for layer $l{+}1$ with computation of layer $l$, with SSD-backed experts passing through the streaming window on their way up. When execution returns to decode, the anchors are released and the borrowed VRAM regions become available for expert caching again.

\paragraph{Decode path.}
When transitioning to decode (green arrows), \sys{} reconfigures the pool into the \emph{Decode Layout}, reclaiming the borrowed VRAM regions as fixed-address expert slots and managing host RAM as an \emph{LRU Cache}. Across this unified pool, experts may reside exclusively in VRAM, exclusively in RAM, or in both tiers. For each layer, the Decode Graphs \textcircled{\scriptsize 1} resolve the routed experts (e.g., $E_A$ and $E_B$) and query the \emph{Residency Table}, which \textcircled{\scriptsize 2} locates their weights across the memory hierarchy. In Figure~\ref{fig:overview}, $E_A$ hits directly in VRAM slot~\#2, whereas $E_B$ misses in both VRAM and RAM. To accommodate $E_B$ in VRAM slot~\#7, the Transfer Engine \textcircled{\scriptsize 3} concurrently demotes the VRAM-only victim~$C$ to RAM while fetching $E_B$ from the SSD into RAM. Once $E_B$ arrives in RAM, the engine \textcircled{\scriptsize 4} promotes it into VRAM slot~\#7, leaving $E_B$ resident in both tiers. Finally, the Memory Manager \textcircled{\scriptsize 5} returns the resolved VRAM slot indices (\#2 and \#7) to the Execution Engine, allowing the \emph{Decode Graphs} to replay captured kernels directly over those fixed slot addresses.

Together, these two paths preserve expert residency across phases. Prefill streams weights without flushing anchored experts in host RAM, while decode demotes evicted VRAM experts back to RAM rather than discarding them, minimizing subsequent SSD reloads. Realizing this architecture requires addressing two system challenges: managing cross-tier residency and layout transitions without wasting memory pool capacity (\S\ref{sec:memory_manager}), and orchestrating asynchronous weight streaming and CUDA Graph replay while expert locations change at runtime (\S\ref{sec:execution_engine}).

\section{Cross-tier Memory Manager}
\label{sec:memory_manager}

The core objective of the Memory Manager is to maximize the effective capacity of VRAM and host RAM across inference phases to avoid costly SSD transfers. As shown in Figure~\ref{fig:overview}, it achieves this through four co-designed mechanisms: a unified \emph{Residency Table} (\S\ref{sec:unified_pool}), \emph{Phase Layout} management (\S\ref{sec:phase_layout}), replica-aware replacement (\S\ref{sec:replica_replacement}), and an asynchronous \emph{Transfer Engine} (\S\ref{sec:expert_transfers}).

\subsection{Unified Memory Pool}
\label{sec:unified_pool}

Conventional offloading systems treat the memory tiers as a cache hierarchy, with VRAM caching RAM and RAM caching the SSD, and let each tier replace its contents independently. On a consumer PC, however, the cost of crossing the tiers is sharply asymmetric. Bidirectional host-to-device (H2D) and device-to-host (D2H) transfers over PCIe complete several times faster than a direct read from the NVMe SSD. An expert resident in RAM remains a valuable fast copy that can be re-served to the GPU at low latency. Likewise, when the GPU evicts an expert that has no host copy, preserving that victim in RAM via a D2H writeback prevents a costly SSD reload on future accesses.

Consequently, rather than managing VRAM and RAM as independent caches, \sys unites them into a Unified Memory Pool.
Within this pool, \sys reserves space for persistent model states and the KV cache, partitioning the remaining VRAM and host RAM into uniform expert slots. The pool's organization adapts to the available capacity. When host RAM alone fits all experts, it holds the complete working copy while VRAM caches active ones. When only the combined VRAM and RAM capacity can fit the model, \sys distributes experts across both tiers and deduplicates them to expand effective capacity. When the combined tiers fall short, experts are paged from the SSD on demand, shifting the manager's primary focus to preventing avoidable pool evictions. To coordinate residency across the pool, the Residency Table maintains a unified directory mapping experts to their VRAM and RAM slots, distinguishing cross-tier duplicates from an expert's last in-memory copy to guide layout adaptation and replacement.

\subsection{Phase-aware Memory Layouts}
\label{sec:phase_layout}

Agent workflows alternate between prefill and decode, which favor different layout patterns in the same memory pool. Prefill scans a broad set of experts and requires space for weight streaming and activations, whereas decode accesses few experts per step and benefits from retaining its evolving working set near the GPU. \sys thus maintains phase-specific layouts that balance execution workspace against expert retention.

During prefill, the manager partitions available VRAM into expert storage, double-buffered weight staging, and activation workspace. In RAM, \emph{anchors} protect reusable experts from being flushed and evicted by prefill scans (\S\ref{sec:eval_residency}), while leaving a small \emph{Streaming Window} to cycle SSD-backed weights. When decode resumes, the manager releases the anchors and reclaims the borrowed VRAM buffers back into fixed-address expert cache slots. Prefill thus uses VRAM for streaming throughput and decode reclaims it for cache capacity, while the anchored RAM-resident set survives across phases to serve subsequent decode requests (\S\ref{sec:replica_replacement}).

\subsection{Replica-aware Replacement}
\label{sec:replica_replacement}

Standard per-tier LRU replacement wastes capacity in a multi-tier pool for two reasons: promoting an expert from the SSD through RAM into VRAM leaves duplicated copies in both tiers, while independent eviction inside each tier is blind to cross-tier redundancy and may evict an expert's last copy in the pool. \sys therefore replaces independent per-tier eviction with a replica-aware policy that coordinates RAM reclamation and GPU victim preservation.

\begin{figure}[t]
    \centering
    \includegraphics[width=0.98\linewidth]{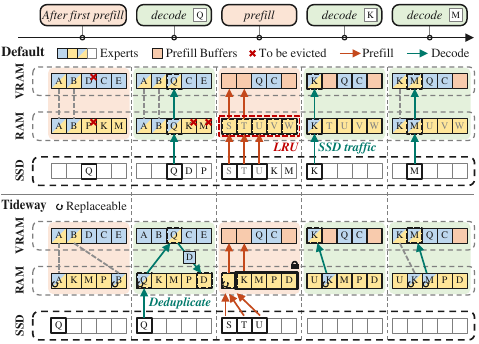}
    \vspace{-3mm}

    \caption{Comparison of cross-tier expert management between default per-tier LRU and \sys across phase transitions.}
    \label{fig:cross_tier_memory}
    \vspace{-3mm}
\end{figure}

\paragraph{Lazy duplicate reclamation.}
When an expert is promoted from RAM to VRAM, \sys retains the RAM copy and marks it as \emph{replaceable} (circular arrows in Figure~\ref{fig:cross_tier_memory}). For RAM allocations, the manager uses free slots first, then reclaims eligible \emph{replaceable} replicas, and finally falls back to LRU among other eligible residents while VRAM evicts by recency as usual. Once either copy is reclaimed, the surviving copy is no longer redundant. Meanwhile, to better balance unique pool capacity against repeated PCIe traffic, \sys exempts a small set of hot experts from duplicate reclamation, allowing them to remain dual-resident.

\paragraph{Selective victim writeback.}
VRAM is the executing tier and evicts far more often than RAM. Under deduplication, many of its victims are VRAM-only experts. Overwriting such a slot drops the expert's last in-pool copy and turns its next access into an SSD read. \sys looks one step further: before eviction, \sys checks whether a valid RAM copy remains. If so, it releases the GPU copy. If host RAM has a free or replaceable slot, \sys demotes the VRAM-only victim back to RAM over the almost idle D2H PCIe link. If neither is available, the victim simply leaves the pool without displacing another RAM-only expert.

Figure~\ref{fig:cross_tier_memory} illustrates how these policies cooperate with the phase-aware layouts. After the first prefill, both Default and \sys hold 8 distinct experts across 10 total slots, with $A$ and $B$ duplicated in both tiers. When decode ~$Q$ requests an SSD-only expert $Q$ and evicts $D$ from VRAM, Default's independent LRU drops $D$ from VRAM and evicts $P$ from RAM to hold $Q$, leaving $Q$ duplicated and shrinking pool coverage to 7 distinct experts. In contrast, \sys reclaims the replaceable RAM copies of $A$ and $B$ while keeping their VRAM copies active, expanding coverage to 9 distinct experts with zero pool evictions. During the subsequent prefill, Default streams incoming weights ($S$--$W$) through its entire RAM under LRU, evicting the in-pool copies of $K$ and $M$. \sys instead locks its unique RAM-only residents ($K, M, P, D$) with anchors and reuses $Q$'s replaceable RAM slot as the streaming window to cycle $S$--$U$. When decode resumes for $K$ and $M$, Default incurs costly SSD traffic to reload both weights, whereas \sys serves both hits directly from host RAM.

\subsection{Asynchronous Expert Transfers}
\label{sec:expert_transfers}

The Transfer Engine turns residency and layout decisions into cross-tier transfers. During prefill, it assembles each layer's weights in the double buffer. VRAM-resident experts are copied within the GPU, RAM-resident experts are transferred over PCIe, and SSD-only experts are loaded through the streaming window. Where possible, contiguous RAM slots are copied together in larger asynchronous transfers, reducing the overhead of issuing separate copies.

During decode, the engine transfers requested experts into their assigned VRAM slots and carries out selective victim writebacks. It uses asynchronous H2D copies to upload RAM-resident weights and
D2H copies to preserve GPU-only victims.
When multiple routed experts miss in the pool, the engine dispatches parallel SSD reads to overlap storage access latencies.
For layouts that retain all experts at fixed locations in the pool, GPU kernels instead read selected weights directly from RAM into VRAM slots, avoiding per-expert CPU dispatch overhead.

As replacement and phase transitions reuse the same storage, \sys coordinates slot reuse with asynchronous transfer completion.
The prefill buffers use two CUDA events. \emph{Release} marks the end of computation using a buffer, and \emph{ready} marks completion of its refill. The copy stream waits for release before loading new weights, while the compute stream waits for ready before using them. These dependencies prevent premature reuse while allowing the execution engine to overlap independent transfers and computation.

\section{Paging-Aware Execution Engine}
\label{sec:execution_engine}

While the Memory Manager determines where experts reside and how memory is partitioned, the \emph{Execution Engine} schedules kernel execution and cross-tier transfers to hide I/O latency and minimize host dispatch overhead. It pairs a \emph{Prefill Pipeline} that overlaps layer-wise weight streaming with computation (\S\ref{sec:prefill_pipeline}), and \emph{Decode Graphs} that capture slot-addressed kernels (\S\ref{sec:graph_capture}) to sustain graph replay under expert paging (\S\ref{sec:graph_replay}).

\subsection{Streaming Prefill Pipeline}
\label{sec:prefill_pipeline}

The Prefill Pipeline processes each input chunk layer by layer over the double buffer provided by the Phase Layout Controller. Once the first layer's weights are staged, the engine submits its computation and asks the Transfer Engine to assemble the next layer in the other buffer. Delivery of layer $l{+}1$ thus runs concurrently with computation of layer $l$, and because a large prefill chunk makes each layer compute-intensive, the computation is long enough to cover the cross-tier transfers behind it.

The double buffer alternates as execution advances: the half consumed by layer $l$ is refilled for layer $l{+}2$, and expert kernels always read their assigned half regardless of whether the weights came from VRAM, RAM, or SSD. To keep the slowest path off the critical path, the engine prepares layer $l{+}1$ after submitting the computation of layer $l$, including any required SSD reads and PCIe staging, and allows its delivery to overlap with the GPU work already in flight. When the last layer completes, the pipeline moves its persistent outputs out of the temporary workspace and returns the borrowed VRAM to the Memory Manager for decoding.

\subsection{Piecewise CUDA Graph Capture}
\label{sec:graph_capture}

When expert paging requires host involvement, a decoder layer cannot be captured as a single monolithic CUDA Graph. Once the router computes active expert IDs on the GPU, the CPU needs to read back those IDs, query the Residency Table, assign target VRAM slots, and issue any required SSD reads and PCIe transfers. As these data-dependent host operations cannot reside inside a captured graph, default offloading engines abandon graph capture along the paged path and fall back to eager kernel dispatch, issuing a long sequence of individual kernel launches per layer. While \sys uses whole-step graphs when the entire expert pool fits in RAM, SSD-backed paging requires synchronizing routing decisions with the host. To retain graph execution without capturing separate graphs for different expert combinations, \sys decomposes each layer into reusable \emph{piecewise CUDA Graphs}, isolating dynamic host paging and variable-width operations.

Under host paging, each layer is split into three segments. The first segment captures the attention prefix such as projections, normalization, and rotary embedding. Sparse attention stays eager. Capturing it would require padding the candidate set to a fixed width, and a different width can change the model's behavior. The second segment captures the front of the FFN, routing weights and expert IDs together with the shared expert, and ends where the host gets the expert routing. The third segment captures the back of the FFN, where routed experts are computed over slot IDs.

\sys{} captures these three segments for each layer once during the warmup phase, binding their inputs and outputs to persistent GPU buffers at fixed addresses. In particular, the third segment reads the normalized inputs, routing weights, and expert IDs directly from the buffers populated by the second segment, avoiding extra data copies across the paging boundary. As the attention segment updates the KV cache and internal states in place, warmup validation snapshots these states before running the reference check and restores them afterward so that capture leaves the model state intact. Once all layers are validated, the captured graphs are sealed, and subsequent tokens simply update the buffer contents in place and replay the segments in layer order without recapturing on expert misses.

\subsection{Paging-Resilient Graph Replay}
\label{sec:graph_replay}

With the segments captured, the remaining question is how to execute when the routed experts may be anywhere in the hierarchy. \sys handles paging on the host between the segments, resolving routed experts into VRAM slots and fetching missing weights as needed. GPU transfers, slot-ID updates, and expert-graph replay are queued in order on the same CUDA stream.
The host can therefore queue subsequent work before transfers finish, while the GPU executes expert kernels only after their weights and indices are ready. Every transfer in the flow is conditional on residency, so a layer whose experts all hit pays only the resolution and the upload, and a miss that reaches the SSD lengthens the handler without changing the execution path around it.

\begin{figure}[t]
    \centering
    \includegraphics[width=\linewidth]{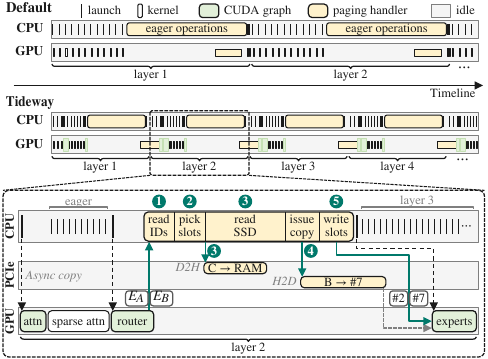}
    \vspace{-8mm}
    \caption{Decode execution timeline under default per-expert eager execution and \sys, and a zoomed-in view of layer~2 in \sys  illustrating piecewise CUDA Graphs (attn, router, experts) coordinated with the five-step paging handler.}
    \label{fig:exec_engine}
    \vspace{-3mm}
\end{figure}

Figure~\ref{fig:exec_engine} expands layer~2 to show this flow in action. The host \textcircled{\scriptsize 1} reads back the expert IDs, and the Memory Manager \textcircled{\scriptsize 2} resolves their residency and assigns VRAM slots. $E_A$ hits in slot~\#2, whereas $E_B$ is absent from the pool and is assigned slot~\#7, replacing GPU-only expert $C$. To prepare this slot, the Transfer Engine \textcircled{\scriptsize 3} overlaps $C$'s asynchronous D2H writeback with the SSD read of $E_B$ into RAM. Once the read completes, it \textcircled{\scriptsize 4} submits the H2D copy of $E_B$ into slot~\#7. Finally, the paging handler \textcircled{\scriptsize 5} uploads the resolved slot IDs (\#2 and \#7) and queues the \texttt{experts} graph behind the transfers on the same stream. The host can then begin queuing layer~3 while the GPU completes layer~2.

\section{Evaluation}
\label{sec:evaluation}

We implement \sys in Python and C, with GPU kernels written in Triton and CuTe DSL.
The implementation comprises about 11.8K source lines, including 5.6K for memory management, 3.8K for execution and kernel adaptations, and 2.3K for backend integration and runtime support.

\paragraph{Experimental setup.}
All of our experiments run on a consumer desktop with a 32-core AMD CPU, 251\,GiB RAM, a 32\,GB NVIDIA GeForce RTX~5090 over PCIe~5.0, and an 8\,TB WD\_BLACK SN850X NVMe SSD, under Linux with NVIDIA driver 580.126, CUDA~13.0, PyTorch~2.11, and Triton~3.6. Measured pinned PCIe bandwidth reaches 57.6\,GB/s H2D, 57.3\,GB/s D2H, and 102.1\,GB/s bidirectional, while SSD reads reach 6.8\,GB/s. We evaluate \textbf{Gemma4-26B-A4B}~\cite{gemmateam2026gemma4} in BF16, \textbf{Qwen3.6-35B-A3B}~\cite{qwen2026qwen36} in BF16, and \textbf{DeepSeek-V4-Flash}~\cite{deepseekai2026deepseekv4} 0731 release in native FP4, capping host RAM at 16--128\,GiB per model to vary memory pressure.

\paragraph{Baselines.}
We compare \sys against four baselines. \textbf{FreeToken}~\cite{yang2026freetoken} v0.1.2 assumes full RAM residency, which we extend with demand SSD paging when weights exceed host RAM. \textbf{llama.cpp}~\cite{gerganov2023llamacpp} commit \texttt{67a17c1} runs with its native CPU--GPU execution and CUDA Graph support. As neither \textbf{SSD-LLaMA}~\cite{liang2026ssdllama} nor \textbf{FlashMoE}~\cite{kim2026flashmoe} is open-source, we reproduce their mechanisms on the unmodified FreeToken backend, implementing SSD-LLaMA's lossless expert compression with CUDA decompression in its all-GPU compute mode and FlashMoE's independently trained per-layer replacement predictors. All systems use the same models and RAM budgets and are tuned with author-recommended configurations.

\paragraph{Workloads and metrics.}
We evaluate \emph{static shapes} with 4K--128K input tokens and 128--2K output tokens, covering 270 configurations across three models and their RAM budgets.
We also replay eight multi-turn coding-agent sessions from \emph{TraceLab}~\cite{zhu2026tracelab}, with growing contexts and repeated prefill--decode transitions.
We report end-to-end completion time, time-to-first-token (TTFT), and decode throughput (tokens/s), averaged within each model--RAM setting, with speedups relative to the fastest baseline.
Unless otherwise stated, prefill uses 8K-token chunks.

\begin{figure*}[t]
  \centering
  \includegraphics[width=\linewidth]{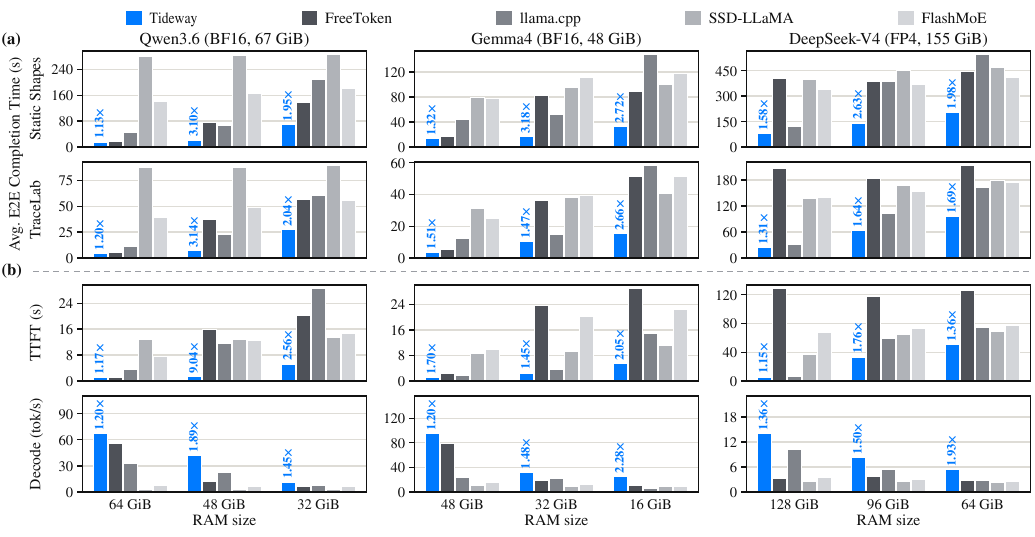}
  \vspace{-8mm}
  \caption{End-to-end inference performance on an RTX 5090 under three host-RAM budgets per model. (a) Average end-to-end completion time per request on Static Shapes and TraceLab. (b) Average time to first token (TTFT) and decode throughput on TraceLab. The number above each Tideway bar is its speedup over the fastest baseline.}
  \label{fig:full-sweep}
  \vspace{-3mm}
\end{figure*}

\subsection{End-to-End Performance}
\label{sec:eval_e2e}

\begin{figure}[t]
    \centering
    \includegraphics[width=\linewidth]{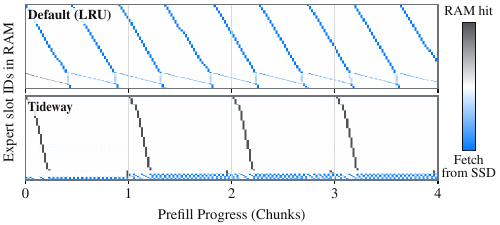}
    \vspace{-8mm}
    \caption{RAM slot residency while prefilling a 32K-token prompt in 8K-token chunks under default LRU and \sys on DeepSeek-V4-Flash at 64\,GiB.}
    \label{fig:prefll_pattern}
    \vspace{-3mm}
\end{figure}

\paragraph{End-to-end completion time.}
Figure~\ref{fig:full-sweep}(a) compares average end-to-end request completion time across all three models and host-RAM budgets.
Across all model-RAM configurations, \sys outperforms every baseline, achieving $1.13\times$--$3.18\times$ speedups ($1.97\times$ on average) on Static Shapes and $1.20\times$--$3.14\times$ on TraceLab.
At the highest RAM budgets, where all experts fit in the unified memory pool, \sys still delivers $1.13\times$--$1.58\times$ speedups by eliminating per-expert CPU dispatch overhead and overlapping PCIe transfers.
\sys's advantage becomes even more pronounced at the lower RAM budgets, where model weights unavoidably spill to the SSD. In these regimes, baselines suffer from severe SSD thrashing as independent per-tier caches retain duplicate copies and flush reusable experts during prefill. \sys minimizes costly SSD reloads, keeps CUDA Graph execution, and hides the remaining transfer latency, widening its speedup to $1.95\times$--$3.18\times$ on Static Shapes and $1.47\times$--$3.14\times$ on TraceLab.

\begin{figure}[t]
    \centering
    \includegraphics[width=\linewidth]{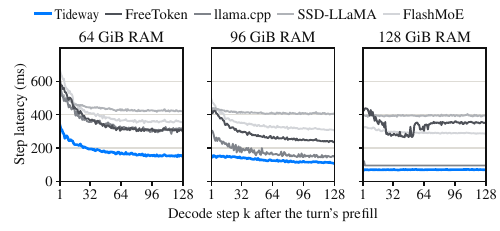}
    \vspace{-8mm}
    \caption{Per-step decode latency after a turn's prefill on DeepSeek-V4-Flash under three RAM budgets, averaged over TraceLab turns.}
    \label{fig:decode_pattern}
    \vspace{-3mm}
\end{figure}

\paragraph{Prefill TTFT.}
As shown in the top row of Figure~\ref{fig:full-sweep}(b), \sys reduces TTFT by $1.15\times$--$9.04\times$ across settings, with a $2.40\times$ average speedup over the fastest baseline.
This improvement comes from two sources. The phase-aware layout (\S\ref{sec:phase_layout}) keeps the RAM-resident experts anchored across chunks, reducing repeated SSD fetches (Figure~\ref{fig:pd}(a)). The Streaming Prefill Pipeline further overlaps the remaining weight transfers with layer-wise GPU computation, using a specialized double buffer and kernels to reduce GPU stalls during weight transportation (\S\ref{sec:prefill_pipeline}).

\paragraph{Decode throughput.}
\sys also accelerates the decode phase.
As shown in the bottom row of Figure~\ref{fig:full-sweep}(b), \sys achieves $1.20\times$--$2.28\times$ higher decode throughput ($1.61\times$ on average) than the fastest baseline.
Phase-aware layouts and replica-aware replacement jointly reclaim prefill workspace and redundant-copy storage, expanding the effective expert capacity of VRAM and RAM and reducing expert I/O (\S\ref{sec:phase_layout}--\S\ref{sec:replica_replacement}).
When paging occurs, the execution engine retains CUDA Graph replay around the paging handler, reducing per-token kernel-launch and CPU orchestration overhead (\S\ref{sec:graph_replay}), as shown in Figure~\ref{fig:pd}(b).

\begin{figure}[t]
    \centering
    \includegraphics[width=\linewidth]{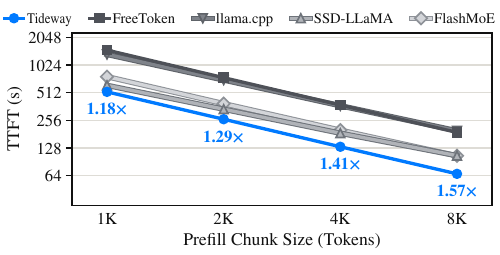}
    \vspace{-8mm}

    \caption{Prefill TTFT across chunk sizes (1K--8K tokens) for a 32K-token prompt on DeepSeek-V4-Flash  with 64\,GiB RAM.}
    \label{fig:chunk_sweep}
    \vspace{-3mm}
\end{figure}

\subsection{Memory Residency under Phase Transitions}
\label{sec:eval_residency}

To understand why \sys sustains high efficiency across phases, we first inspect host-RAM slot accesses during a 32K-token prefill with 8K-token prefill chunks on DeepSeek-V4 with 64\,GiB RAM in Figure~\ref{fig:prefll_pattern}.
Under default per-tier LRU, streaming a full pass of layers exceeds host RAM capacity and overwrites existing slots in sequence, producing diagonal blue sweeps across the entire RAM pool. As each layer sweep evicts the experts needed by the next sweep, almost every slot access across all four chunks degrades into an SSD fetch. In contrast, \sys{} exhibits recurring RAM hits across chunks, with SSD fetches concentrated in a small streaming window. Anchors preserve reusable residents while the window absorbs the scan traffic (\S\ref{sec:phase_layout}), allowing successive chunks to reuse cached weights.

Figure~\ref{fig:decode_pattern} further examines how prefill affects subsequent decoding by tracking per-step latency after each turn's prefill. Across all three RAM budgets, \sys starts at a lower initial latency and converges more quickly to a lower steady-state latency than every baseline. Even at 128\,GiB RAM, where the model fits in memory, several baselines still suffer a sharp cold-start spike and remain elevated or exhibit step-to-step fluctuations. These observations are consistent with reduced disruption to expert residency during prefill, limiting the recovery cost when decode resumes.

\subsection{Sensitivity Analysis on Prefill Chunk Size}
\label{sec:eval_sensitivity}

Prefill chunk size decides how often the expert pool is swept per prompt, so we evaluate its effect on TTFT for DeepSeek-V4-Flash with a 32K-token input in Figure~\ref{fig:chunk_sweep}. Increasing the chunk size from 1K to 8K steadily reduces TTFT across all systems.
Since each chunk accesses nearly all experts even at small chunk sizes, larger chunks require fewer model-wide weight sweeps, amortizing transfer costs over more tokens.
In our experiments, we cap the prefill chunk size at 8K to limit activation-memory growth and preserve space for double buffering and resident experts on the GPU.

\sys achieves the lowest TTFT throughout, with its speedup over the fastest baseline increasing from $1.18\times$ at 1K to $1.57\times$ at 8K.
Anchored RAM residency reduces repeated SSD reads across chunks (\S\ref{sec:phase_layout}).
As the chunk size increases toward 8K, each layer becomes progressively more compute-intensive, providing the Streaming Prefill Pipeline (\S\ref{sec:prefill_pipeline}) with a longer execution window per layer to hide lookahead SSD reads and double-buffered PCIe transfers behind GPU computation.
In contrast, the baselines remain bottlenecked by unoverlapped I/O stalls and repeated cache thrashing across passes, causing \sys's relative advantage to widen at larger chunk sizes.

\subsection{Ablation Study}
\label{sec:eval_ablation}

\begin{figure}[t]
    \centering
    \includegraphics[width=\linewidth]{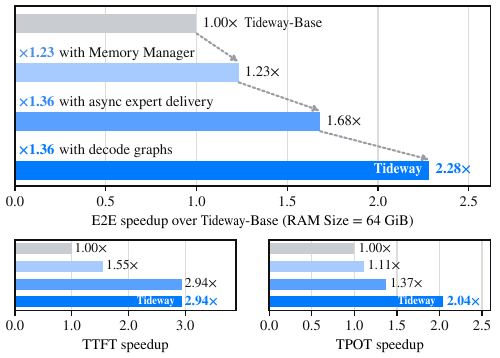}
    \vspace{-8mm}
    \caption{Ablation study of \sys on DeepSeek-V4-Flash with 32K-token input, 1,024-token output, and 8K prefill chunks.}
    \label{fig:ablation}
    \vspace{-3mm}
\end{figure}

To quantify the contribution of each proposed mechanism, Figure~\ref{fig:ablation} presents a factor-by-factor breakdown on DeepSeek-V4-Flash with a 32K-token input, a 1K-token output, and 8K prefill chunks. We start from \sys-Base, which uses per-tier LRU eviction, synchronous expert loading, and eager execution. Enabling the Cross-tier Memory Manager in \S\ref{sec:memory_manager} protects reusable RAM residents during prefill and reclaims cross-tier duplicates, yielding a $1.23\times$ end-to-end speedup. Adding asynchronous expert delivery and prefill pipelining from \S\ref{sec:expert_transfers} and \S\ref{sec:prefill_pipeline} overlaps SSD reads and double-buffered PCIe transfers with GPU computation, providing a further $1.36\times$ gain. Finally, introducing piecewise decode graphs from \S\ref{sec:graph_capture}--\S\ref{sec:graph_replay} removes host dispatch overhead under expert paging, delivering another $1.36\times$ gain and bringing the overall end-to-end speedup to $2.28\times$.

The bottom panels of Figure~\ref{fig:ablation} show the three mechanisms' speedup in different phases. For TTFT, the Memory Manager raises it to $1.55\times$ by cutting SSD fetches, and asynchronous delivery to $2.94\times$ by overlapping transfers and computation. Decode graphs bring no speedup to TTFT. For TPOT, the first two mechanisms lift it to $1.11\times$ and $1.37\times$, and the bulk of the $2.04\times$ comes from decode graphs reducing host dispatch stalls under expert paging.

\section{Discussion}
\label{sec:discussion}

\paragraph{Runtime overhead and SSD endurance.}
\sys's mechanisms cost little at run time. Capturing and validating the piecewise CUDA Graphs happens once during warmup and completes in a few seconds, and the persistent buffers and Residency Table take under 3\% of the 32\,GB of VRAM. All writeback operations, including selective D2H victim writebacks, are between VRAM and host RAM over PCIe, with the NVMe drive acting as a read-only backing store. As the unified pool and scan protection cut the bytes read from SSD per request (\S\ref{sec:motivation:moe}), \sys also reduces storage I/O pressure on consumer drives.

\paragraph{KV cache as a competing tenant.}
\sys reserves KV capacity before allocating expert slots, so growing contexts leave less memory for expert residency.
Extending the pool to manage KV blocks alongside experts could combine host offloading and prefix reuse~\cite{zheng2024sglang} to balance memory pressure against recomputation.
Sparse KV access~\cite{tang2024quest,lee2024infinigen} could further reduce the transferred data.
WiSP~\cite{zhang2026wisp} and MARS~\cite{wang2026mars} show that the joint expert and KV working set bounds effective capacity on low-resource hardware, requiring admission control over this shared budget when serving concurrent requests.

\paragraph{Generality and orthogonal optimizations.}
Our design is orthogonal to weight quantization, speculative decoding, and lookahead expert prefetching. As shown in \S\ref{sec:evaluation}, \sys supports both BF16 and native FP4 quantized models, and can be combined with lossless storage compression when PCIe outpaces SSD bandwidth. Similarly, multi-token verification in speculative decoding maps naturally onto \sys{}'s slot-addressed decode graphs by widening the fixed token dimension of the captured segments without recapturing on page misses.

\section{Related Work}
\label{sec:related_work}

\paragraph{Offloading-based LLM inference.}
Early offloading engines such as FlexGen~\cite{sheng2023flexgen} and DeepSpeed-Inference~\cite{aminabadi2022deepspeed} rely on static layer-wise schedules tailored to offline batch throughput. For sparse MoE models, Fiddler~\cite{kamahori2024fiddler} and KTransformers~\cite{ktransformers2024} execute cold experts on the host CPU to save PCIe bandwidth, yet remain bottlenecked by host compute. Conversely, GPU-centric MoE systems including MoE-Infinity~\cite{xue2024moeinfinity} and SwapMoE~\cite{kong2023swapmoe} overlap PCIe transfers via lookahead prefetching and activation-guided caching. While these designs assume sufficient host DRAM and fixed cache hierarchies, \sys manages VRAM and RAM as a unified pool to sustain execution under expert paging.

\paragraph{Expanding and sharing GPU memory.}
Recent systems also explore expanding VRAM capacity and supporting multiple applications on a single GPU. CUDA Unified Memory~\cite{nvidiaUVM} oversubscribes VRAM via page-granular demand paging. TGS~\cite{wu2023tgs}, nvshare~\cite{alexopoulos2024nvshare}, and Nixie~\cite{xu2026nixie} enable multiple applications to share a single GPU, treating each workload's allocations as opaque memory blocks. \sys operates within a single model at expert granularity. Integrating \sys with these system-level multiplexers remains an interesting future direction.

\paragraph{Agentic workflows and graph execution.}
ReAct-style agents interleave generation with external feedback across turns~\cite{yao2023react}, repeatedly invoking prefill and decode as contexts grow.
Trending frontends such as Codex CLI~\cite{openai2026codexsubagents} and OpenCode~\cite{opencode2026agents} further support spawning subagents, creating request dependencies at runtime.
Parrot~\cite{lin2024parrot} exposes cross-request dataflow for application-level optimization, while Autellix~\cite{luo2025autellix} schedules calls using program-level progress.
At the GPU level, FlashInfer~\cite{ye2025flashinfer} combines dynamic attention scheduling with CUDA Graph replay.
Integrating workflow-aware scheduling with \sys could coordinate subagent concurrency and expert residency when multiple sessions share a local model.

\section{Conclusion}
\label{sec:conclusion}

Frontier MoE models are increasingly deployed to power local agents.
We present \sys{}, a phase-aware tiered memory management and execution system for efficient agentic inference on
consumer PCs. \sys unites GPU VRAM and host RAM into a unified memory pool, combining phase-specific layouts with replica-aware replacement and selective victim writeback to maximize distinct expert coverage and prevent prefill scan thrashing. Paired with a streaming prefill pipeline and piecewise CUDA Graph replay under expert paging, \sys{} achieves $1.97\times$ average end-to-end speedup on static workloads and up to $3.14\times$ speedup on multi-turn agent traces over state-of-the-art baselines.

\section{Acknowledgments}
\label{sec:ack}
This work was supported in part by the National Science Foundation (NSF) (Grant No. 2625580) and the Army Research Office (ARO) (Grant No. W911NF-23-2-0224).

{\small
\bibliographystyle{mlsys2025}
\bibliography{references}

@inproceedings{yao2023react,
  author    = {Shunyu Yao and Jeffrey Zhao and Dian Yu and Nan Du and
               Izhak Shafran and Karthik Narasimhan and Yuan Cao},
  title     = {{ReAct}: Synergizing Reasoning and Acting in Language Models},
  booktitle = {The Eleventh International Conference on Learning Representations
               (ICLR 2023)},
  year      = {2023}
}

@inproceedings{xie2024osworld,
  author    = {Tianbao Xie and Danyang Zhang and Jixuan Chen and
               Xiaochuan Li and Siheng Zhao and Ruisheng Cao and
               Toh Jing Hua and Zhoujun Cheng and Dongchan Shin and
               Fangyu Lei and Yitao Liu and Yiheng Xu and Shuyan Zhou and
               Silvio Savarese and Caiming Xiong and Victor Zhong and Tao Yu},
  title     = {{OSWorld}: Benchmarking Multimodal Agents for Open-Ended Tasks
               in Real Computer Environments},
  booktitle = {Advances in Neural Information Processing Systems
               (NeurIPS 2024)},
  volume    = {37},
  year      = {2024}
}

@misc{openaiDeepResearch2025,
  author    = {{OpenAI}},
  title     = {Introducing {Deep Research}},
  year      = {2025},
  month     = feb,
  url       = {https://openai.com/index/introducing-deep-research/},
  note      = {Accessed: 2026-10-05}
}

@misc{anthropic2025claudecode,
  author    = {{Anthropic}},
  title     = {{Claude 3.7 Sonnet} and {Claude Code}},
  year      = {2025},
  month     = feb,
  url       = {https://www.anthropic.com/news/claude-3-7-sonnet},
  note      = {Accessed: 2026-10-05}
}

@misc{openai2025codex,
  author    = {{OpenAI}},
  title     = {Introducing {Codex}},
  year      = {2025},
  month     = may,
  url       = {https://openai.com/index/introducing-codex/},
  note      = {Accessed: 2026-10-05}
}

@article{yang2025qwen3,
  author    = {An Yang and Anfeng Li and Baosong Yang and Beichen Zhang and
               Binyuan Hui and Bo Zheng and Bowen Yu and Chang Gao and
               Chengen Huang and Chenxu Lv and Chujie Zheng and
               Dayiheng Liu and Fan Zhou and Fei Huang and Feng Hu and
               Hao Ge and Haoran Wei and Huan Lin and Jialong Tang and
               Jian Yang and Jianhong Tu and Jianwei Zhang and
               Jianxin Yang and Jiaxi Yang and Jing Zhou and Jingren Zhou and
               Junyang Lin and Kai Dang and Keqin Bao and Kexin Yang and
               Le Yu and Lianghao Deng and Mei Li and Mingfeng Xue and
               Mingze Li and Pei Zhang and Peng Wang and Qin Zhu and
               Rui Men and Ruize Gao and Shixuan Liu and Shuang Luo and
               Tianhao Li and Tianyi Tang and Wenbiao Yin and
               Xingzhang Ren and Xinyu Wang and Xinyu Zhang and
               Xuancheng Ren and Yang Fan and Yang Su and Yichang Zhang and
               Yinger Zhang and Yu Wan and Yuqiong Liu and Zekun Wang and
               Zeyu Cui and Zhenru Zhang and Zhipeng Zhou and Zihan Qiu},
  title     = {{Qwen3} Technical Report},
  journal   = {arXiv preprint arXiv:2505.09388},
  year      = {2025}
}

@article{deepseekai2026deepseekv4,
  author    = {{DeepSeek-AI}},
  title     = {{DeepSeek-V4}: Towards Highly Efficient Million-Token
               Context Intelligence},
  journal   = {arXiv preprint arXiv:2606.19348},
  year      = {2026}
}

@article{gemmateam2026gemma4,
  author    = {{Gemma Team}},
  title     = {{Gemma 4} Technical Report},
  journal   = {arXiv preprint arXiv:2607.02770},
  year      = {2026}
}

@inproceedings{song2024powerinfer,
  author    = {Yixin Song and Zeyu Mi and Haotong Xie and Haibo Chen},
  title     = {{PowerInfer}: Fast Large Language Model Serving with a
               Consumer-grade {GPU}},
  booktitle = {Proceedings of the ACM SIGOPS 30th Symposium on Operating Systems
               Principles (SOSP 2024)},
  pages     = {590--606},
  year      = {2024}
}

@inproceedings{erdogan2024tinyagent,
  author    = {Lutfi Eren Erdogan and Nicholas Lee and Siddharth Jha and
               Sehoon Kim and Ryan Tabrizi and Suhong Moon and
               Coleman Richard Charles Hooper and Gopala Anumanchipalli and
               Kurt Keutzer and Amir Gholami},
  title     = {{TinyAgent}: Function Calling at the Edge},
  booktitle = {Proceedings of the 2024 Conference on Empirical Methods in
               Natural Language Processing: System Demonstrations
               (EMNLP 2024)},
  pages     = {80--88},
  year      = {2024}
}

@misc{trendforce2026memory,
  author    = {{TrendForce}},
  title     = {{AI} Server Demand Sustains Memory Contract Price Increases
               in {4Q26}, While Consumer-Side Pressure Persists,
               Says {TrendForce}},
  year      = {2026},
  month     = sep,
  url       = {https://www.trendforce.com/presscenter/news/20260930-13258.html},
  note      = {Accessed: 2026-10-05}
}

@article{yang2026freetoken,
  author    = {Shuo Yang and Xiaoze Fan and Melissa Pan and Haocheng Xi and
               Zhe Wang and Shanlin Sun and Kurt Keutzer and Song Han and
               Matei Zaharia and Chenfeng Xu and Ion Stoica},
  title     = {{FreeToken}: Efficient Edge-Native {MoE} Serving with
               Bandwidth-Adaptive Execution},
  journal   = {arXiv preprint arXiv:2608.16157},
  year      = {2026}
}

@misc{gerganov2023llamacpp,
  author    = {Georgi Gerganov and others},
  title     = {{llama.cpp}: {LLM} Inference in {C/C++}},
  year      = {2023},
  url       = {https://github.com/ggml-org/llama.cpp},
  note      = {Accessed: 2026-10-05}
}

@article{kim2026flashmoe,
  author    = {Byeongju Kim and Jungwan Lee and Donghyeon Han and
               Hoi-Jun Yoo and Sangyeob Kim},
  title     = {{FlashMoE}: Reducing {SSD I/O} Bottlenecks via {ML}-Based
               Cache Replacement for Mixture-of-Experts Inference
               on Edge Devices},
  journal   = {arXiv preprint arXiv:2601.17063},
  year      = {2026}
}

@article{liang2026ssdllama,
  author    = {Fangzhou Liang and Yibin Shen and Jianmin Hu and
               Jiayang Xu and Hanchi Gao and Minxian Xu and Zili Meng},
  title     = {{SSD-LLaMA}: {SSD}-Native Inference for Trillion-Parameter
               {MoE} at 1+ Token/s on a Consumer {PC}},
  journal   = {arXiv preprint arXiv:2609.18110},
  year      = {2026}
}

@misc{ul2026threedmark,
  author       = {{UL Solutions}},
  title        = {{3DMark} Benchmark Results Database},
  howpublished = {\url{https://www.3dmark.com/search}},
  year         = {2026},
  note         = {Submitted results with an {RTX}~5090, Q3 2026. Accessed: 2026-10-09}
}

@misc{boinc2026hosts,
  author       = {{BOINC}},
  title        = {{BOINC} project host statistics},
  howpublished = {\url{https://boinc.berkeley.edu/}},
  year         = {2026},
  note         = {Host records with an RTX~5090 active in 2026, collected from the public host lists of PrimeGrid, MilkyWay@home, and Einstein@Home. Accessed: 2026-10-09}
}

@misc{valve2026steam,
  author       = {{Valve Corporation}},
  title        = {Steam Hardware \& Software Survey},
  howpublished = {\url{https://store.steampowered.com/hwsurvey/}},
  year         = {2026},
  note         = {Q3 2026 results. Accessed: 2026-10-09}
}

@misc{qwen2026qwen36,
  author       = {{Qwen Team}},
  title        = {{Qwen3.6-35B-A3B}},
  howpublished = {Hugging Face model card, \url{https://huggingface.co/Qwen/Qwen3.6-35B-A3B}},
  year         = {2026},
  month        = apr,
  note         = {Accessed: 2026-10-10}
}

@article{zhu2026tracelab,
  author  = {Kan Zhu and Mathew Jacob and Chenxi Ma and Yi Pan and
             Stephanie Wang and Arvind Krishnamurthy and Baris Kasikci},
  title   = {{TraceLab}: Characterizing Coding Agent Workloads for {LLM} Serving},
  journal = {arXiv preprint arXiv:2606.30560},
  year    = {2026}
}

@inproceedings{tang2024quest,
  author    = {Jiaming Tang and Yilong Zhao and Kan Zhu and Guangxuan Xiao and Baris Kasikci and Song Han},
  title     = {Quest: Query-Aware Sparsity for Efficient Long-Context {LLM} Inference},
  booktitle = {Proceedings of the 41st International Conference on Machine Learning (ICML 2024)},
  year      = {2024}
}

@inproceedings{lee2024infinigen,
  author    = {Wonbeom Lee and Jungi Lee and Junghwan Seo and Jaewoong Sim},
  title     = {{InfiniGen}: Efficient Generative Inference of Large Language Models with Dynamic {KV} Cache Management},
  booktitle = {18th USENIX Symposium on Operating Systems Design and Implementation (OSDI 2024)},
  year      = {2024}
}

@inproceedings{sheng2023flexgen,
  author    = {Ying Sheng and Lianmin Zheng and Binhang Yuan and
               Zhuohan Li and Max Ryabinin and Beidi Chen and
               Percy Liang and Christopher R{\'e} and Ion Stoica and
               Ce Zhang},
  title     = {{FlexGen}: High-Throughput Generative Inference of Large
               Language Models with a Single {GPU}},
  booktitle = {Proceedings of the 40th International Conference on Machine Learning
               (ICML 2023)},
  series    = {Proceedings of Machine Learning Research},
  volume    = {202},
  pages     = {31094--31116},
  year      = {2023}
}

@inproceedings{zheng2024sglang,
  author    = {Lianmin Zheng and Liangsheng Yin and Zhiqiang Xie and
               Chuyue Sun and Jeff Huang and Cody Hao Yu and
               Shiyi Cao and Christos Kozyrakis and Ion Stoica and
               Joseph E. Gonzalez and Clark Barrett and Ying Sheng},
  title     = {{SGLang}: Efficient Execution of Structured Language
               Model Programs},
  booktitle = {Advances in Neural Information Processing Systems
               (NeurIPS 2024)},
  volume    = {37},
  pages     = {62557--62583},
  year      = {2024}
}

@inproceedings{kamahori2024fiddler,
  author    = {Keisuke Kamahori and Tian Tang and Yile Gu and Kan Zhu and Baris Kasikci},
  title     = {{Fiddler}: {CPU-GPU} Orchestration for Fast Inference of Mixture-of-Experts Models},
  booktitle = {Proceedings of the 13th International Conference on Learning Representations (ICLR 2025)},
  year      = {2025}
}

@misc{ktransformers2024,
  author       = {{KVCache.AI Team}},
  title        = {{KTransformers}: A Flexible Framework for Experiencing Cutting-edge {LLM} Inference Optimizations},
  howpublished = {\url{https://github.com/kvcache-ai/ktransformers}},
  year         = {2024},
  note         = {Accessed: 2026-10-05}
}

@article{kong2023swapmoe,
  author    = {Rui Kong and Yuanchun Li and Weijun Wang and Linghe Kong and Yunxin Liu},
  title     = {Serving {MoE} Models on Resource-Constrained Edge Devices via Dynamic Expert Swapping},
  journal   = {IEEE Transactions on Computers},
  volume    = {74},
  number    = {8},
  pages     = {2799--2811},
  year      = {2025}
}

@inproceedings{aminabadi2022deepspeed,
  author    = {Reza Yazdani Aminabadi and Samyam Rajbhandari and Ammar Ahmad Awan and
               Cheng Li and Du Li and Elton Zheng and Olatunji Ruwase and
               Shaden Smith and Minjia Zhang and Jeff Rasley and Yuxiong He},
  title     = {{DeepSpeed-Inference}: Enabling Efficient Inference of Transformer Models at Unprecedented Scale},
  booktitle = {Proceedings of the International Conference for High Performance Computing, Networking, Storage and Analysis (SC 2022)},
  pages     = {1--15},
  year      = {2022}
}

@article{xue2024moeinfinity,
  author    = {Leyang Xue and Yao Fu and Zhan Lu and Luo Mai and Mahesh Marina},
  title     = {{MoE-Infinity}: Offloading-Efficient {MoE} Model Serving},
  journal   = {arXiv preprint arXiv:2401.14361},
  year      = {2024}
}

@article{wang2026mars,
  author    = {Yifei Wang and Hancheng Ye and Yechen Xu and
               Cong Guo and Chiyue Wei and Qinsi Wang and
               Dongting Li and Tingjun Chen and Hai {``Helen''} Li and
               Danyang Zhuo and Yiran Chen},
  title     = {{MARS}: Efficient, Adaptive Co-Scheduling for
               Heterogeneous Agentic Systems},
  journal   = {arXiv preprint arXiv:2604.26963},
  year      = {2026}
}

@misc{nvidiaUVM,
  author       = {{NVIDIA}},
  title        = {{CUDA} {C++} Programming Guide: Unified Memory Programming},
  howpublished = {\url{https://docs.nvidia.com/cuda/cuda-c-programming-guide/}},
  year         = {2026},
  note         = {Accessed: 2026-10-24}
}

@inproceedings{wu2023tgs,
  author    = {Bingyang Wu and Zili Zhang and Zhihao Bai and Xuanzhe Liu and Xin Jin},
  title     = {Transparent {GPU} Sharing in Container Clouds for Deep Learning Workloads},
  booktitle = {20th USENIX Symposium on Networked Systems Design and Implementation (NSDI 2023)},
  pages     = {69--85},
  year      = {2023}
}

@inproceedings{alexopoulos2024nvshare,
  author    = {Georgios Alexopoulos and Dimitris Mitropoulos},
  title     = {nvshare: Practical {GPU} Sharing without Memory Size Constraints},
  booktitle = {Proceedings of the 46th IEEE/ACM International Conference on Software Engineering: Companion Proceedings (ICSE 2024)},
  year      = {2024}
}

@inproceedings{xu2026nixie,
  author    = {Yechen Xu and Yifei Wang and Nathanael Ren and Yiran Chen and Danyang Zhuo},
  title     = {Nixie: Efficient, Transparent Temporal Multiplexing for Consumer {GPUs}},
  booktitle = {20th USENIX Symposium on Operating Systems Design and Implementation (OSDI 2026)},
  pages     = {2085--2101},
  year      = {2026}
}

@misc{openai2026codexsubagents,
  author       = {{OpenAI}},
  title        = {{Codex} Documentation: Subagents},
  howpublished = {\url{https://developers.openai.com/codex/subagents}},
  year         = {2026},
  note         = {Accessed: 2026-10-24}
}

@misc{opencode2026agents,
  author       = {{OpenCode}},
  title        = {{OpenCode} Documentation: Agents},
  howpublished = {\url{https://opencode.ai/docs/agents/}},
  year         = {2026},
  note         = {Accessed: 2026-10-24}
}

@inproceedings{lin2024parrot,
  author    = {Chaofan Lin and Zhenhua Han and Chengruidong Zhang and Yuqing Yang and Fan Yang and Chen Chen and Lili Qiu},
  title     = {Parrot: Efficient Serving of {LLM}-based Applications with Semantic Variable},
  booktitle = {18th USENIX Symposium on Operating Systems Design and Implementation (OSDI 2024)},
  year      = {2024}
}

@article{luo2025autellix,
  author    = {Michael Luo and Xiaoxiang Shi and Colin Cai and Tianjun Zhang and Justin Wong and Yichuan Wang and Chi Wang and Yanping Huang and Zhifeng Chen and Joseph E. Gonzalez and Ion Stoica},
  title     = {Autellix: An Efficient Serving Engine for {LLM} Agents as General Programs},
  journal   = {arXiv preprint arXiv:2502.13965},
  year      = {2025}
}

@inproceedings{ye2025flashinfer,
  author    = {Zihao Ye and Lequn Chen and Ruihang Lai and Wuwei Lin and Yineng Zhang and Stephanie Wang and Tianqi Chen and Baris Kasikci and Vinod Grover and Arvind Krishnamurthy and Luis Ceze},
  title     = {{FlashInfer}: Efficient and Customizable Attention Engine for {LLM} Inference Serving},
  booktitle = {Proceedings of the 8th Conference on Machine Learning and Systems (MLSys 2025)},
  year      = {2025}
}

@article{zhang2026wisp,
  author    = {Jiamu Zhang and Liang Wu and Mayank Darbari and Liangjie Hong},
  title     = {{WiSP}: A Working-Set View of Mixture-of-Experts Serving on Extremely Low-Resource Hardware},
  journal   = {arXiv preprint arXiv:2606.21868},
  year      = {2026},
  doi       = {10.48550/arXiv.2606.21868},
  url       = {https://arxiv.org/abs/2606.21868}
}
}

\end{document}